# mlGW: a machine-learning framework for accurate and scalable *GW* quasiparticle calculations

Fan Zhang[1+], Weiwei Gao[2,3*+], Jijun Zhao[4,5*], James R. Chelikowsky[6,7,8], Fengqi Song[1,2,3]

[1]National Laboratory of Solid State Microstructures, Collaborative Innovation Center of Advanced Microstructures, and School of Physics, Nanjing University, Nanjing 210093, China.

[2]Institute of Atom Manufacturing, Nanjing University, Nanjing 210089, China.

[3]Jiangsu Provincial Key Laboratory of Atomic Level Manufacturing, Nanjing Institute of Atomic Scale Manufacturing, Nanjing 211800, China

[4]Guangdong Provincial Key Laboratory of Quantum Engineering and Quantum Materials, School of Physics, South China Normal University, Guangzhou 510006, China

[5]Guangdong-Hong Kong Joint Laboratory of Quantum Matter, Frontier Research Institute for Physics, South China Normal University, Guangzhou 510006, China

[6]Center for Computational Materials, Oden Institute for Computational Engineering and Sciences, The University of Texas at Austin, Austin, Texas 78712, United States of America

[7]Department of Physics, The University of Texas at Austin, Austin, Texas 78712, United States of America

[8]McKetta Department of Chemical Engineering, The University of Texas at Austin, Austin, Texas 78712, United States of America

## Abstract

Many-body perturbation methods such as the *GW* approximation yield quasiparticle energies beyond the reach of conventional density functional theory (DFT). Although recent algorithmic developments have extended *GW* calculations to tens of thousands of atoms, such calculations typically require leadership-class computing resources. We developed a machine-learning *GW* (mlGW) framework that uses a graph neural network to predict state-resolved *GW* corrections from a compact descriptor of the target Kohn-Sham state and its local electronic environment. These descriptors can be constructed either from explicit DFT calculations or from machine-learned Hamiltonians, the latter enabling quasiparticle-energy predictions without explicit DFT calculations on the target system. mlGW accurately predicts quasiparticle energies across closed-shell molecules, crystals, point defects, and Si nanoclusters, and captures the electron–phonon-induced band-gap renormalization of diamond and silicon. Except for the localized defect states, all these applications reuse a single trained model without task-specific fine-tuning. The computational cost of mlGW inference scales empirically as $O(N)$ with system size, while the predicted quasiparticle energies achieve mean absolute errors of 15.4 and 12.6 meV on QM9 and the tested crystals, respectively. Combined with a machine-learned Hamiltonian, mlGW enables quasiparticle electronic-structure predictions for a hydrogen-passivated silicon nanocluster containing 30,115 atoms (109,768 valence electrons) at a modest computational cost.

Machine learning (ML) has become a powerful paradigm for accelerating first-principles calculations of molecules and condensed matter, enabling efficient prediction of energies, forces, electronic Hamiltonians, and other properties[1-9]. Most of these ML models, however, are trained on data generated at the level of density functional theory (DFT). While DFT offers a favorable balance between accuracy and computational cost[10], it is formally a ground-state theory: Kohn-Sham eigenvalues lack a rigorous interpretation as quasiparticle energies, and fundamental band gaps and other excited-state properties are often poorly described. ML models trained on DFT data inherit these limitations, and quantitatively accurate quasiparticle energies call for many-body treatments[11,12].

The *GW* approximation provides accurate quasiparticle band structures and serves as the starting point for electron-phonon coupling and electron-hole interaction calculations[11-14]. Extending ML to *GW* quasiparticle energies, however, is challenging because the quasiparticle self-energy is nonlocal and energy-dependent and because generating *GW* reference data for training is computationally expensive. Recent developments in efficient, highly parallelized *GW* algorithms[15-25] have alleviated this data bottleneck, making it practical to generate large training datasets for surrogate models. Pioneering studies have encoded mean-field electronic information in physically motivated descriptors or learned latent representations of Kohn-Sham wavefunctions[26-31], while more general approaches learn many-body self-energies or Green's functions directly from mean-field features[32] or atomic structures[33]. Together, these works establish the feasibility of predicting many-body corrections from mean-field inputs. However, existing ML-based *GW* approaches have been benchmarked mainly on either molecules and small clusters[26,32,34,35] or crystalline materials[27-31], and their accuracy and scalability have not been demonstrated for localized defect states or for systems beyond a few thousand atoms[28].

Here, we introduce a machine-learning *GW* (mlGW) framework for accurate and scalable quasiparticle calculations across molecules, clusters, crystals, and point defects. mlGW encodes atomic species, the orbital weight of the target Kohn-Sham state, and $\boldsymbol{k}$-resolved occupied orbital populations into a compact descriptor, from which a graph neural network (GNN) predicts state-resolved *GW* corrections. A key feature is that these descriptors can be constructed either from explicit DFT calculations or from a machine-learned Hamiltonian. We demonstrate the accuracy and transferability of the mlGW model by computing the temperature-dependent electron-phonon-induced band-gap renormalization of diamond and silicon and by predicting the spin-resolved defect states of the nitrogen-vacancy ($NV^{-}$) center in diamond. Combined with a machine-learned Hamiltonian, mlGW further extends quasiparticle calculations to hydrogen-passivated Si nanoclusters containing more than 30,000 atoms, at a computational cost several orders of magnitude lower than that of conventional DFT.

## Results

### The general architecture of the mlGW model

We work with the $G_0W_0$ approximation, in which the self-energy is constructed once from the DFT Kohn-Sham states and evaluated at the quasiparticle energy within the linearized quasiparticle equation. The $GW$ correction is given by $\Delta E_{n\boldsymbol{k}}^{GW} = \langle\psi_{n\boldsymbol{k}}|\hat{\Sigma}^{\mathrm{GW}}(E_{n\boldsymbol{k}}^{GW}) - \hat{V}_{xc}|\psi_{n\boldsymbol{k}}\rangle$, with $E_{n\boldsymbol{k}}^{GW} = E_{n\boldsymbol{k}}^{DFT} + \Delta E_{n\boldsymbol{k}}^{GW}$. The self-energy is nonlocal and frequency dependent, and its evaluation requires both the target Kohn-Sham state and the screened Coulomb interaction of the system. Rather than explicitly constructing the dielectric matrix, mlGW learns a mapping from state-resolved electronic-structure descriptors to the $GW$ correction. As shown in Fig. 1, the input embedding $e_{i,n\boldsymbol{k}}^{0}$ of node $i$ combines the atomic species, the state-resolved atomic-orbital coefficient weights $w_{n\boldsymbol{k}}^{il}$, and the $\boldsymbol{k}$-resolved occupied orbital population $q_{\boldsymbol{k}}^{il}$:

$$e_{i,n\boldsymbol{k}}^{0} = [Z_i|w_{n\boldsymbol{k}}^{is}|w_{n\boldsymbol{k}}^{ip}|w_{n\boldsymbol{k}}^{id}|q_{\boldsymbol{k}}^{is}|q_{\boldsymbol{k}}^{ip}|q_{\boldsymbol{k}}^{id}], \quad (1)$$

where $Z_i$ is a learnable embedding of the atomic species of atom $i$. These descriptors can be constructed from explicit DFT calculations or, alternatively, from machine-learned Hamiltonians[4-6,36] in a localized atomic basis. Here we use the double-zeta valence-plus-polarization (DZP) basis set for atomic-orbital projection.

In the explicit DFT route, the self-consistent Kohn-Sham Hamiltonian is represented in a nonorthogonal basis of atom-centered localized orbitals. The atomic-orbital expansion coefficients are obtained by solving the corresponding generalized eigenvalue problem

$$\mathbf{H}(\boldsymbol{k})\mathbf{C}_{n\boldsymbol{k}} = E_{n\boldsymbol{k}}^{AO}\mathbf{S}(\boldsymbol{k})\mathbf{C}_{n\boldsymbol{k}}, \quad (2)$$

where $\mathbf{C}_{n\boldsymbol{k}}$ is the atomic-orbital coefficient vector, and $\boldsymbol{S}(\boldsymbol{k})$ is the overlap matrix in the nonorthogonal localized basis. This eigenvalue problem is solved with

ScaLAPACK[37]. The state-specific orbital weight is defined as

$$w_{n\boldsymbol{k}}^{il} = \sum_{\mu \in \mathcal{A}(i,l)} \left|C_{\mu n \boldsymbol{k}}\right|^2, \qquad l = s, p, d \tag{3}$$

where $\mathcal{A}(i,l)$ denotes the set of atomic orbitals belonging to atom $i$ and angular-momentum channel $l$, and $C_{\mu n\boldsymbol{k}}$ is the expansion coefficient of electronic state $n\boldsymbol{k}$ in orbital $\mu$.

In the machine-learned Hamiltonian route, we use DeepH[4], a previously developed ML model for predicting DFT Hamiltonians from atomic structures, to obtain the Hamiltonian matrix $\mathbf{H}(\boldsymbol{k})$ in the same DZP atomic-orbital basis. The atomic-orbital expansion coefficients $\mathbf{C}_{n\boldsymbol{k}}$ are obtained by solving Eq. (2), and the corresponding state-specific orbital weights $w_{n\boldsymbol{k}}^{il}$ are calculated using Eq. (3). The $\boldsymbol{k}$-resolved occupied orbital population is defined as

$$q_{\boldsymbol{k}}^{il} = \sum_{m=1}^{N_{occ}} w_{m\boldsymbol{k}}^{il}, \tag{4}$$

where $N_{occ}$ is the number of occupied Kohn-Sham states. $q_{\boldsymbol{k}}^{il}$ represents the total occupied orbital weight on the $l$-channel of atom $i$ at wavevector $\boldsymbol{k}$, providing a local descriptor of the occupied electronic environment. Unlike a real-space charge density, $q_{\boldsymbol{k}}^{il}$ is a basis-projected, $\boldsymbol{k}$-resolved reduced quantity. Both $w_{n\boldsymbol{k}}^{il}$ and $q_{\boldsymbol{k}}^{il}$ are real and invariant to the global phase of nondegenerate states.

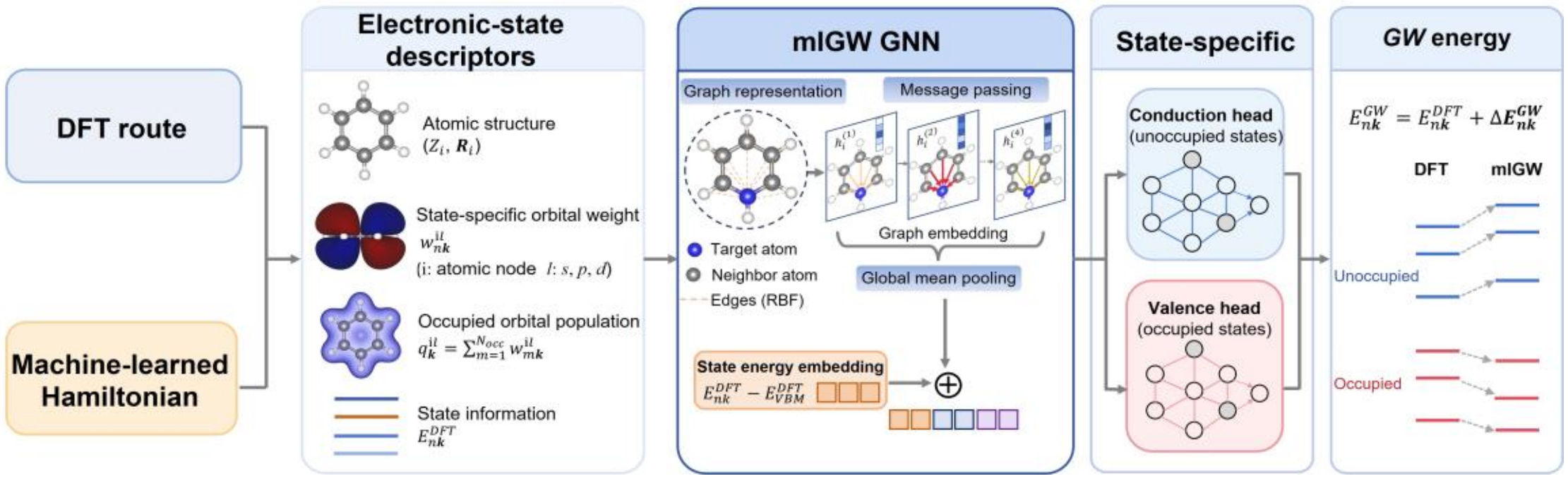

**Fig. 1 Overview of the mlGW framework.** State-resolved electronic descriptors are obtained from either explicit DFT calculations or machine-learned Hamiltonians. A graph neural network encodes the atomic structure and these descriptors into a graph-level embedding, which is concatenated with a state-energy embedding and passed to separate prediction heads for occupied and unoccupied states, yielding the *GW* quasiparticle energies.

The initial node embeddings $e_{i,\boldsymbol{nk}}^{0}$ are processed by a stack of residual message-passing layers[38], followed by global mean pooling to obtain a graph-level embedding. Occupied and unoccupied states receive systematically different *GW* corrections. Thus, this embedding is concatenated with a state-energy embedding and passed to separate prediction heads for the two classes, while sharing the same graph-level representation. The state energy is measured relative to the highest occupied level, $E_{\boldsymbol{nk}}^{DFT} - E_{VBM}^{DFT}$, where $E_{VBM}^{DFT}$ is the VBM for crystals (in case of molecules, $E_{VBM}^{DFT}$ should be replaced by HOMO). For spin-polarized calculations, a discrete spin embedding is additionally concatenated to label each state as non-spin-polarized, spin-up, or spin-down.

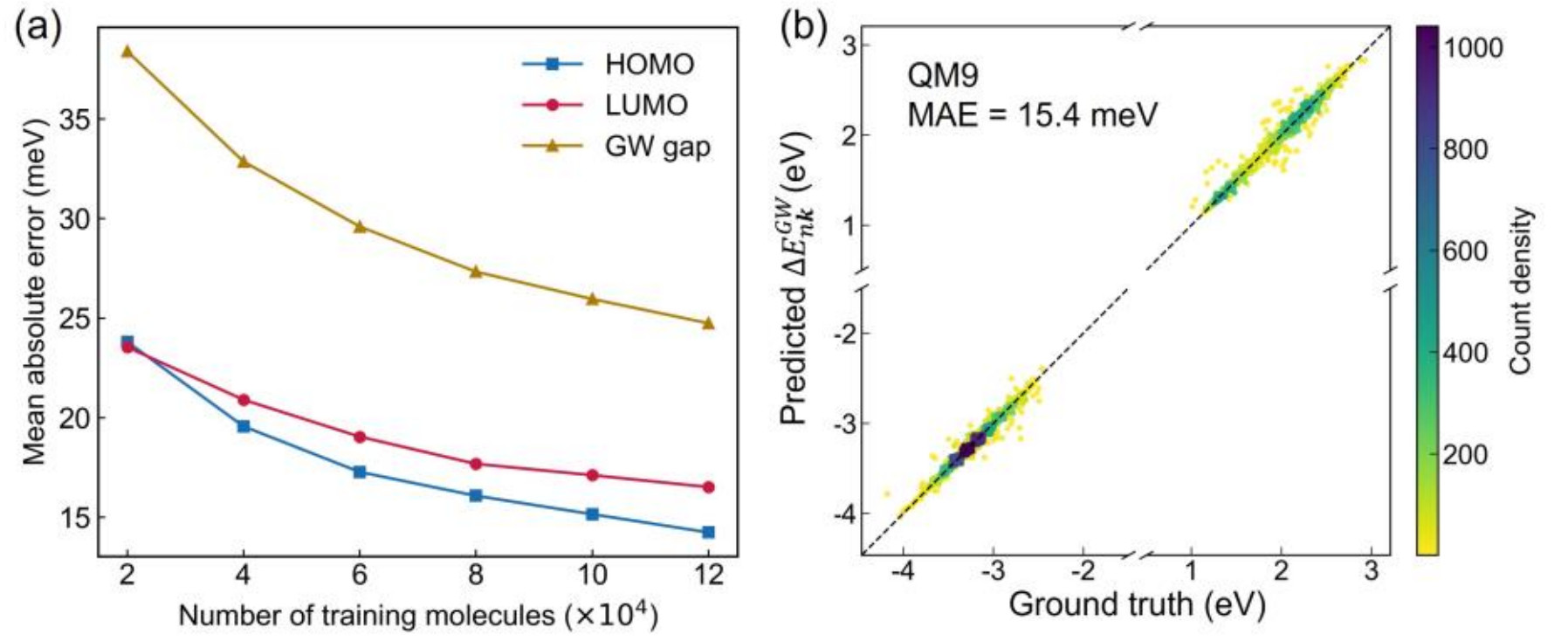

**Fig. 2 Accuracy of mlGW for QM9 molecules.** (a) MAE of the HOMO, LUMO $G_0W_0$ correction and the resulting GW gap in the test set as a function of the number of QM9 training samples. The test set is fixed at 5,000 molecules. (b) Predicted versus reference *GW* corrections of the test set for QM9 molecules. The color scale indicates the density of data points.

To demonstrate the accuracy on closed-shell molecules, we first benchmarked mlGW on the QM9 dataset, which contains 133,885 molecules composed of H, C, N, O, and F, with up to nine non-hydrogen atoms. Reference *GW* quasiparticle energies for the HOMO and LUMO were taken from the dataset reported by Fediai *et al.* [39]. A fixed set of 5,000 molecules was held out for testing, while the remaining molecules were used to construct training sets of various sizes. As a baseline, a linear regression from the DFT eigenvalue to the *GW* correction yields substantial mean absolute errors (MAEs) of 159 and 276 meV for the HOMO and LUMO, respectively (Fig. S1), indicating that the correction cannot be captured by the eigenvalue alone. Using mlGW, the prediction errors decrease systematically with training-set size, reaching 14.3 and 16.5 meV for the HOMO and LUMO corrections when 120,000 molecules are used for training (Fig. 2a). The corresponding predicted-versus-reference quasiparticle corrections for the 5,000-molecule test set are displayed in Fig. 2b, yielding an overall MAE of 15.4 meV. Table I compares mlGW with representative ML approaches for *GW* quasiparticle energies. Because DimeNet++[34] was trained on GW reference data from the same source, this is a like-for-like comparison, and mlGW achieves lower MAEs for both HOMO and LUMO states (and a lower gap MAE of 24.8 meV). The accuracy is comparable to that of MBGF-Net[32]; the key distinction is that mlGW targets transferable and scalable corrections across molecules, nanoclusters, crystals, and defects, as demonstrated below.

**Table I. Comparison of ML approaches for predicting *GW* quasiparticle energies of QM9 molecules.** ViSNet additionally uses DFT-level pretraining data. As these studies employ different training strategies, *GW* reference methods, and test sets, the reported accuracies should be interpreted as a cross-study comparison rather than a direct benchmark.

| Method | MAE of HOMO (meV) | MAE of LUMO (meV) | MAE of GW gap (meV) |
|---|---|---|---|
| DimeNet++[34] | 22 | 31 | 42 |
| MBGF-Net[32] | 17 | 18 | 29 |
| ViSNet[35] | $\approx 50$ | $\approx 35$ | $\approx 50$ |
| mlGW (this work) | 14.3 | 16.5 | 24.8 |

## Accuracy and transferability for crystals

We next applied mlGW to semiconductors and insulators using a dataset of 142 structural configurations containing H, C, N, O, F, and Si, with fewer than 40 atoms per unit cell. The dataset includes elemental substances, $H_2O$- and $NH_3$-based molecular systems, and Si-C, Si-N, Si-O compounds. Crystal structures were obtained from the Materials Project[40,41]. The dataset composition and representative structures are shown in Fig. S2. Reference *GW* calculations were performed using the BerkeleyGW package[42]; computational details are given in **Methods**. At each sampled ***k***-point, we extracted the *GW* corrections for the three highest occupied valence bands and the three lowest unoccupied conduction bands. The resulting dataset contains 29,868 Kohn-Sham states sampled across different bands and ***k***-points, together with their reference quasiparticle energies. The states were split into training and test sets in an 8:2 ratio. On the test set, mlGW achieves MAEs of 6.5 and 18.5 meV for occupied and unoccupied states, respectively, yielding an overall MAE of 12.6 meV (Fig. 3a). To assess transferability beyond the training structures, we applied the trained model to several crystals that are excluded from the training dataset. Figs. 3b, 3c, and 3d compare the DFT, explicit $G_0W_0$, and mlGW band structures of $Si_3N_4$ (mp-11607), Si (mp-1095269), and SiC (mp-7631); additional unseen structures, including C (mp-616440) and $SiO_2$ (mp-556319), are shown in Fig. S3. In all cases, mlGW closely reproduces the state-resolved quasiparticle corrections across both valence and conduction bands, confirming that the model is transferable to crystal structures outside the training set.

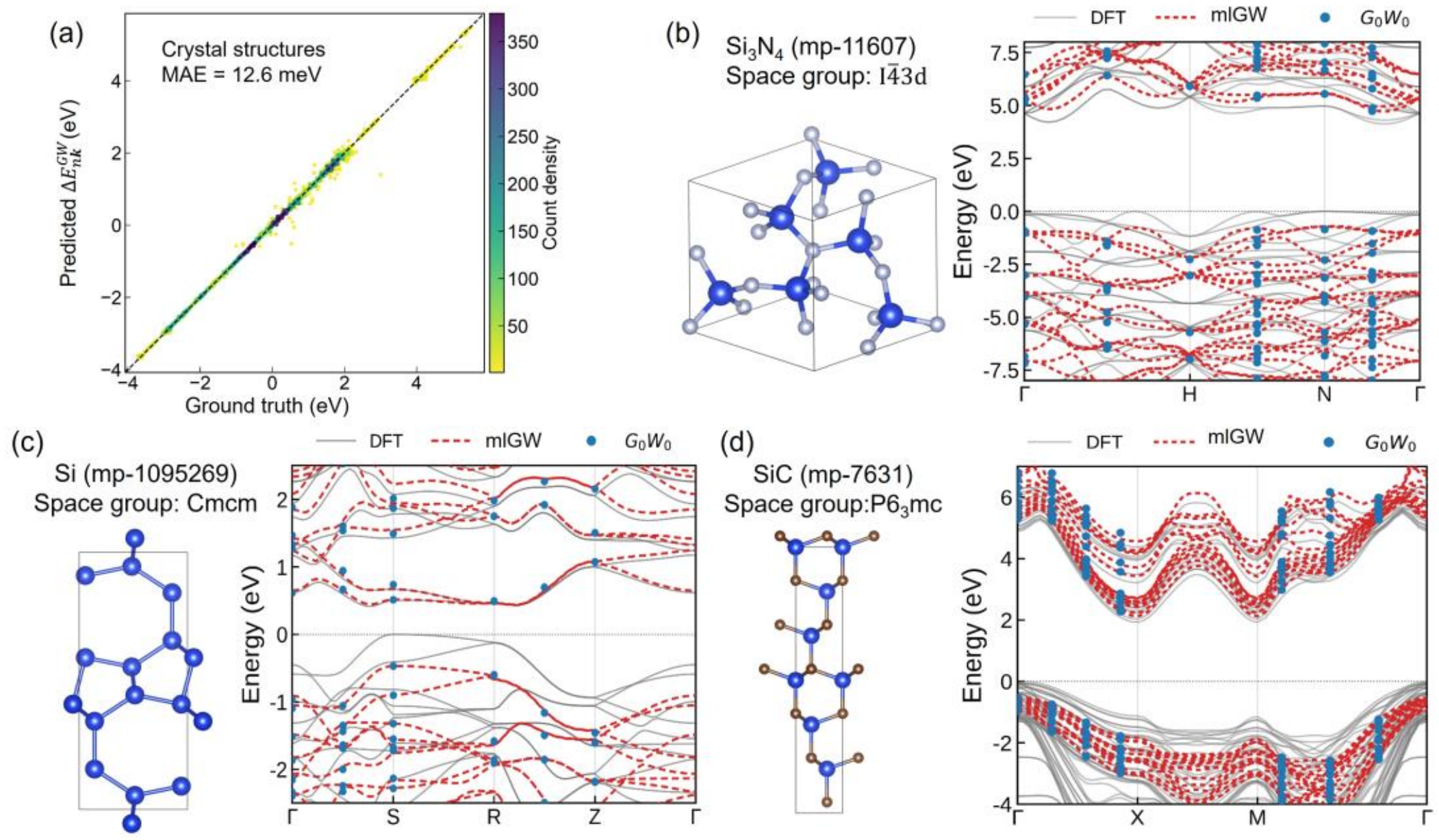

**Fig. 3 Accuracy and transferability of mlGW across crystals.** (a) Predicted versus reference *GW* corrections of test set for crystalline materials. The color scale indicates the density of data points. (b, c, d) Atomic structures and quasiparticle band structures of (b) $Si_3N_4$ (mp-11607), (c) Si (mp-1095269) and (d) SiC (mp-7631), all excluded from the training set. Gray solid lines, red dashed lines, and blue circles represent DFT, mlGW, and explicit $G_0W_0$ results, respectively.

## Electron-phonon-induced band-gap renormalization

Electron-phonon (*e-ph*) coupling strongly affects electronic states and is a major source of the temperature dependence of semiconductor band gaps[43-46]. Many-body correlation effects can further modify the *e-ph*-induced band-gap renormalization in diamond and silicon[47], motivating a quasiparticle-level treatment. We combined material-specific machine-learned Hamiltonians[4] with the mlGW model to compute the temperature-dependent *e-ph* renormalization of the fundamental gaps of diamond and silicon; the mlGW model is exactly the one trained on the crystal dataset above, reused without retraining or fine-tuning. At each temperature, a representative 9×9×9 supercell (1458 atoms) was generated within the harmonic approximation by combining the displacements of all phonon modes, with amplitudes determined by their temperature-dependent occupations[48] (see **Methods**). At $T = 0$ K, the displacements arise solely from zero-point motion, and the zero-point renormalization (ZPR) is obtained as the change of the quasiparticle gap relative to the equilibrium structure. Table II compares the resulting ZPRs with previous theoretical and experimental values: the *GW*-level ZPR of diamond deviates from experiment by about 80 meV, while that of silicon agrees with experiment to within 5 meV. Diamond exhibits a substantially larger ZPR than silicon, consistent with previous calculations[49].

**Table II. ZPRs of the indirect gaps in diamond and silicon.** The $G_0W_0$ values for this work are obtained by combining a machine-learned Hamiltonian and mlGW, while the reference values were taken from experiments as well as conventional DFT and $G_0W_0$ calculations in the literature. All energies are in meV.

| | Diamond | | Silicon | |
|---|---|---|---|---|
| | DFT | $G_0W_0$ | DFT | $G_0W_0$ |
| This work | −356 | −489 | −75 | −66 |
| Theory Reference | −320[43] | −337[43] | −57[48] | -- |
| Experiment Reference | −410[50] | | −62[51] | |

Fig. 4a and 4b show the temperature-induced changes of the fundamental indirect gaps of crystalline diamond and silicon relative to their 0 K values. To isolate the *e-ph* contribution from static gap errors, the calculated and experimental gaps are aligned at 0 K; with this alignment, the mlGW results reproduce the measured temperature dependence more closely than the DFT (PBE) results. The combination of machine-learned Hamiltonians and mlGW thus captures quasiparticle gap renormalization across thermally distorted configurations. Fig. S4 further compares the diamond renormalization obtained from the two descriptor routes, machine-learned Hamiltonian +mlGW and explicit DFT+mlGW. These two agree to within 7.5 meV over the entire temperature range, validating the use of machine-learned Hamiltonians in combination with mlGW.

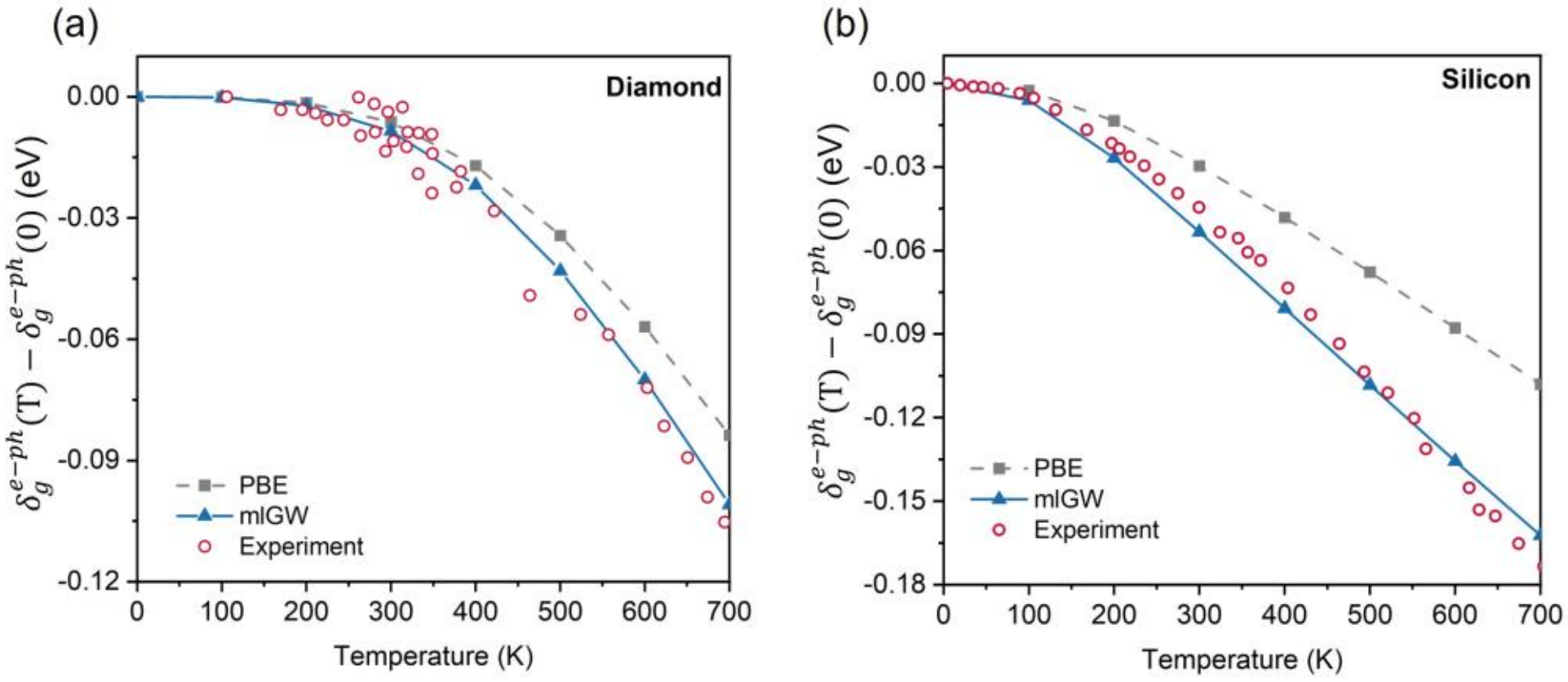

**Fig. 4 Temperature-dependent *e-ph*-induced band-gap renormalization in diamond and silicon.** Changes in the fundamental indirect band gaps relative to their 0 K values are shown for (a) diamond and (b) silicon. Experimental data[52,53] are shown by red open circles, PBE results by gray squares and dashed lines, and mlGW results by blue triangles and solid lines.

## Large-scale quasiparticle calculations for Si nanoclusters

To demonstrate scalability, we applied the framework to hydrogen-passivated Si nanoclusters containing up to 30,115 atoms, with diameters ranging from approximately 1.2 to 10.3 nm. The machine-learned Hamiltonian model was trained on 596 Si-based structures (hydrogen-passivated nanoclusters, slabs, and bulk configurations) with the largest training structure containing only 180 atoms, so that the workflow extrapolates to clusters more than two orders of magnitude larger. Electronic states were obtained from the machine-learned Hamiltonian, and quasiparticle corrections were evaluated with the same crystal-trained mlGW model, reused without retraining. Note that no explicit DFT or *GW* calculations were performed for the target nanoclusters. Fig. 5a shows the size dependence of the band gaps obtained from the machine-learned Hamiltonian model and the mlGW model. Both gaps decrease systematically with increasing diameter and approach their respective bulk limits, reflecting the weakening of quantum confinement. The size dependence can be fitted with a power-law relation

$$E_g(D) = E_g^{bulk} + \frac{\mathrm{A}}{D^{\alpha}}, \quad (5)$$

where $D$ is the cluster diameter, $E_g^{bulk}$ is the band gap of bulk Si, and A and α are fitting parameters. The bulk gaps were fixed at their calculated values of 0.61 and 1.12 eV for DFT and GW, respectively, and only clusters containing 489 atoms or more were included in the fitting. The predicted *GW* gaps coincide well with the size dependence reported from stochastic GW calculations[54] and scanning tunneling spectroscopy measurements on individual Si nanocrystals[55], whereas the DFT gaps systematically underestimate the experimental values by approximately 0.6 eV. For the largest cluster, $Si_{26551}H_{3564}$, the predicted *GW* gap is approximately 1.19 eV, close to the calculated bulk limit of 1.12 eV.

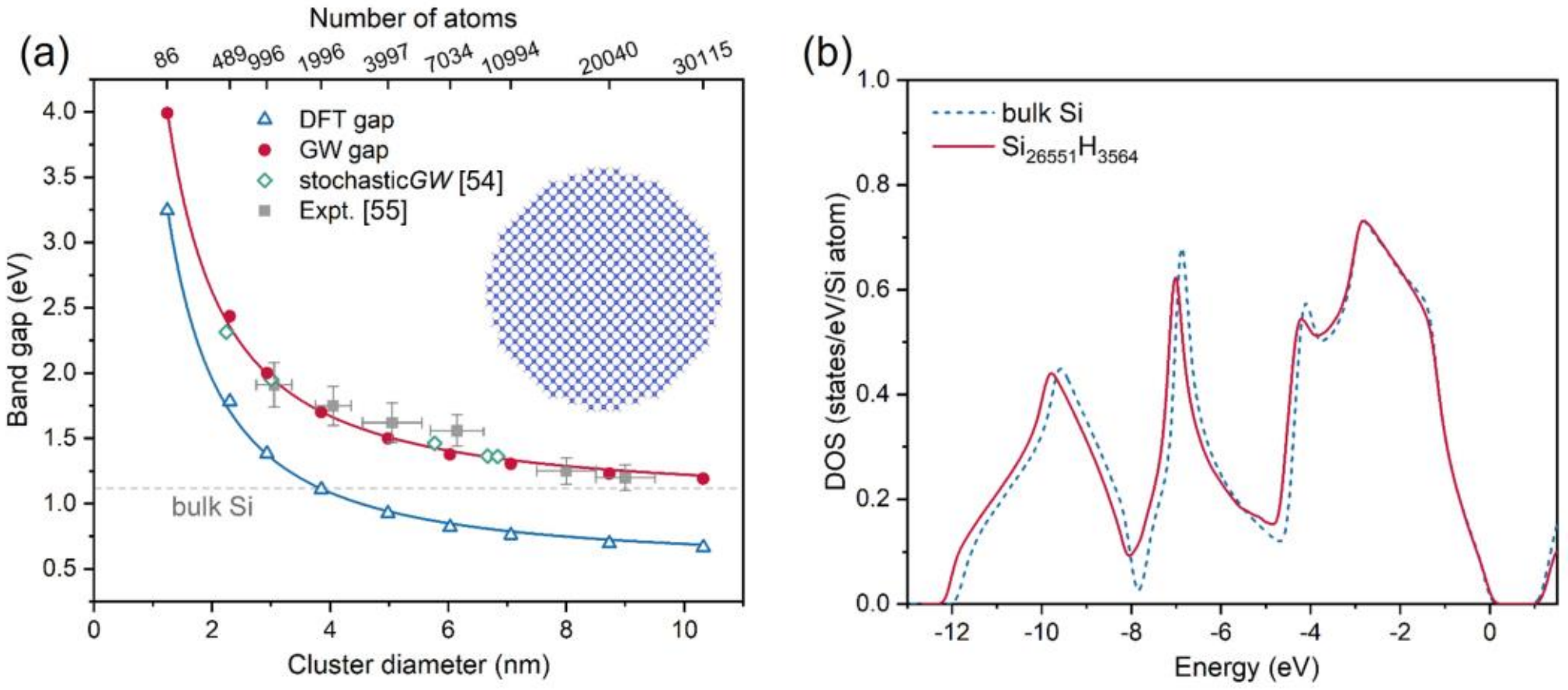

**Fig. 5 Large-scale mlGW calculations for hydrogen-passivated Si nanoclusters.** (a) Size dependence of the band gaps calculated using machine-learned Hamiltonian (blue triangles) and mlGW (red circles), compared with reference stochastic *GW* calculations[54] (green diamonds) and experimental measurements[55] (gray squares). Solid lines represent power-law fits with the corresponding bulk band gaps fixed at their calculated values, restricted to clusters with ≥489 atoms. The horizontal dashed line indicates the bulk Si quasiparticle gap, and the inset shows a representative Si nanocluster. (b) Quasiparticle DOS of bulk Si obtained from explicit *GW* calculations (blue dashed line) and the $Si_{26551}H_{3564}$ nanocluster predicted using mlGW with machine-learned Hamiltonian model (red solid line). The two spectra are aligned at the valence-band edge (HOMO for the cluster), broadened with Gaussians of standard deviation 0.10 eV, and normalized per Si atom so that their intensities are directly comparable.

Beyond the fundamental band gap, we examined the quasiparticle density of states (DOS) of the largest nanocluster. Fig. 5b compares the DOS of $Si_{26551}H_{3564}$, constructed from machine-learned Hamiltonian states with mlGW correction, with the reference bulk Si DOS from explicit GW calculations. The nanocluster DOS reproduces the main features of the bulk quasiparticle spectrum, including the characteristic valence-band peaks, as expected for a cluster of this size approaching the bulk limit. Remaining differences may arise from finite-size effects, surface passivation, and residual errors of the ML models.

Figs. 6a and 6b compare the computational cost and peak memory of Hamiltonian diagonalization, mlGW inference, and conventional DFT calculations, each over the system-size range accessible to it. Over the investigated range, diagonalization scales approximately as $O(N^3)$ in cost and $O(N^2)$ in peak memory, whereas both the cost and memory of mlGW inference scale approximately linearly with $N$ for one state. At comparable system sizes, the combined cost of diagonalization and mlGW inference is several orders of magnitude lower than that of conventional DFT; for system sizes beyond the range of explicit DFT calculations, the DFT cost is estimated by extrapolating the conventional plane-wave DFT scaling (Fig. 6a). For the largest cluster, $Si_{26551}H_{3564}$ (30,115 atoms and 109,768 valence electrons), Hamiltonian diagonalization requires approximately $1.98 \times 10^4$ core-hours and 2.42 TB of peak memory on Hygon C86 7285H processors, while mlGW inference adds only a minor overhead. The bottleneck of the mlGW with machine-learned Hamiltonian workflow is therefore the generalized eigenvalue problem required to produce the electronic-state descriptors, not the prediction of the quasiparticle corrections themselves.

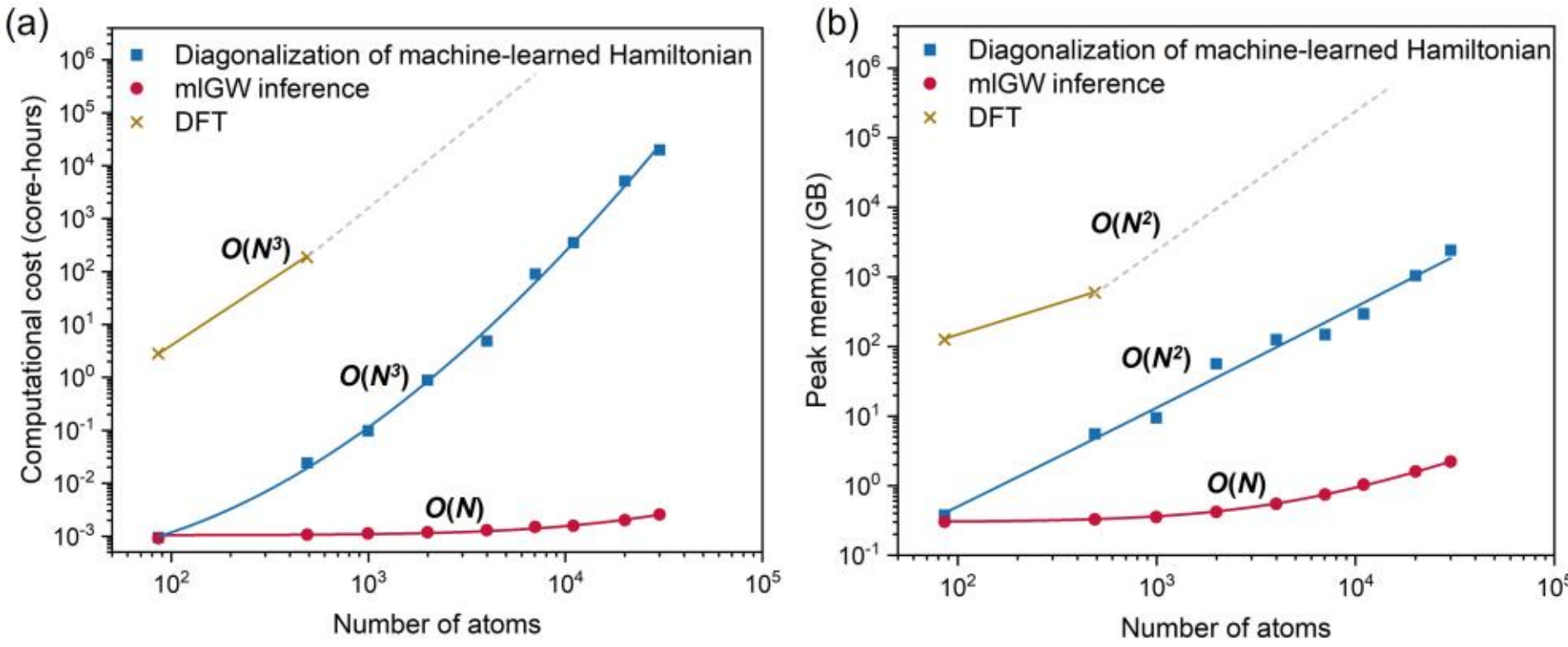

**Fig. 6 Computational cost and peak memory of large-scale mlGW calculations.** (a) Computational cost in core-hours and (b) peak memory usage as functions of system size for diagonalization of machine-learned Hamiltonian (blue squares), mlGW inference (red circles), and DFT calculations (yellow crosses). Solid lines indicate the corresponding scaling trends. Gray dashed lines show extrapolations assuming conventional plane-wave DFT scaling, with computational cost proportional to $N^3$ and memory usage proportional to $N^2$. Computational cost is defined as $N_{core} \times t_{wall}$, where $N_{core}$ is the number of CPU cores and $t_{wall}$ is measured wall time for the task.

## Spin-resolved quasiparticle states of the $NV^-$ center

Finally, we tested mlGW on localized defect states of the $NV^-$ center in diamond[56]. To represent defect-localized, spin-polarized states, we retrained mlGW on the crystalline training set augmented with seven defect-containing diamond supercells of different sizes: one carbon vacancy ($C_{63}$), three substitutional-nitrogen configurations ($C_{15}N$, $C_{31}N$, and $C_{47}N$), and three $NV^-$ centers ($C_{30}N$, $C_{62}N$, and $C_{126}N$). The 215- and 511-atom supercells of $NV^-$ were excluded from the training set to assess transferability to larger defect systems.

Fig. 7a shows the atomic structure and local defect environment of the 215-atom $NV^-$ supercell. Fig. 7b compares the spin-resolved Γ-point levels obtained from DFT, explicit GW (BerkeleyGW), and the explicit-DFT+mlGW route for the 215-atom supercell, and the DFT+mlGW route for the 511-atom supercell. mlGW captures the *GW*-induced shifts of both bulk-like band-edge states and localized defect states, with an MAE of 140 meV across the eight Γ-point states of the 215-atom cell—the bulk-like valence-band maximum (VBM), six spin-resolved defect levels, and the bulk-like conduction-band minimum (CBM).

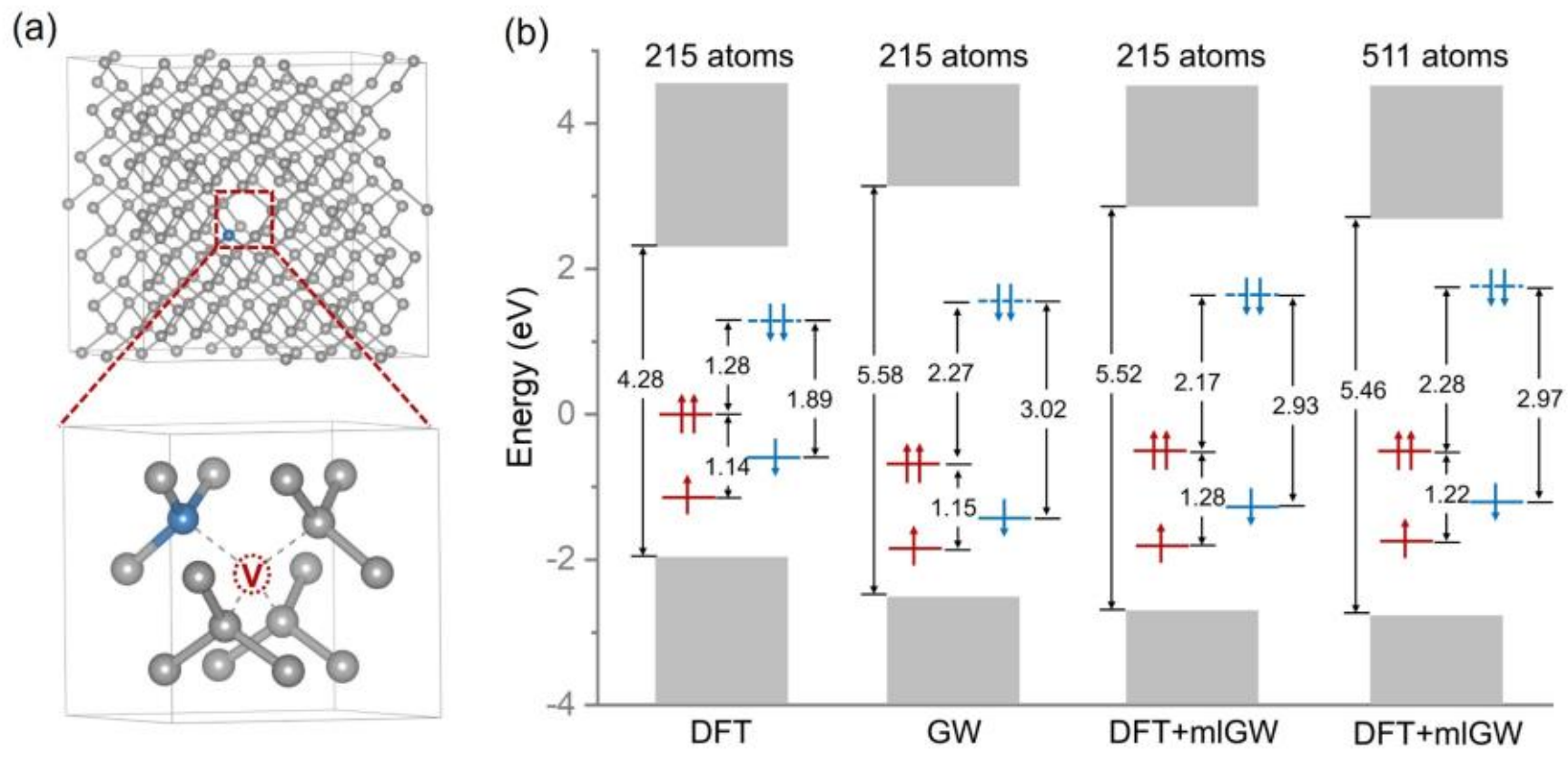

**Fig. 7 Atomic structure and spin-resolved electronic states of the $NV^-$ center in diamond.** (a) A 215-atom diamond supercell containing an $NV^-$ center, with an enlarged view of the local defect structure. Carbon and nitrogen atoms are shown in gray and blue, respectively, and the vacancy is indicated by a red dashed circle. (b) Comparisons of band edges and spin-resolved defect levels for the 215-atom supercell (DFT, GW, and DFT+mlGW) and for the 511-atom supercell (DFT+mlGW only). Gray shaded regions represent the bulk valence and conduction bands, while red and blue arrows indicate spin-up and spin-down states, respectively. The energy zero is set to the highest occupied DFT level, which serves as the common reference for all calculations.

## Discussion

In summary, we have introduced mlGW, a framework for predicting state-resolved quasiparticle corrections, and demonstrated its applications across molecules, nanoclusters, crystals, and point defects. The atom- and orbital-resolved electronic descriptors enable accurate predictions of *GW* quasiparticle energies. These descriptors can be obtained either from explicit DFT calculations or from machine-learned Hamiltonians; the latter route eliminates the need for explicit DFT calculations on the target system.

A closely related approach, DeepH-GW[57], also targets large-scale GW calculations. It directly maps to material structure to the GW Hamiltonian in an localized atomic-orbital basis, from which quasiparticle energies are obtained through diagonalization, and has been demonstrated on large supercells of diamond and GaP crystals. In contrast, mlGW adopts a Δ-learning formulation, learning state-resolved scalar corrections to the Kohn-Sham energies from electronic-state descriptors.

A practical advantage of the mlGW framework is its modular training workflow. The two modules learn different mappings: a machine-learned Hamiltonian maps atomic structures to Hamiltonian matrix elements, whereas mlGW maps an electronic state and its chemical environment to the *GW* correction. They are trained independently and can therefore draw on different data distributions. In the applications above, a single mlGW model trained on the periodic-crystal dataset was reused without fine-tuning or target-specific training data for unseen crystals, for the *e-ph*-induced band-gap renormalization of diamond and silicon, and for hydrogen-passivated Si nanoclusters containing up to 30,115 atoms. The Hamiltonian side, by contrast, was task-specific: separate machine-learned Hamiltonian models were trained for the bulk materials and for the Si nanoclusters. The $NV^-$ center is the exception: for this strongly localized defect case, we augmented the mlGW training set with seven defect-containing supercells. This modularity can improve data efficiency when the state-to-correction mapping transfers better than the structure-to-Hamiltonian mapping, and it allows either module to be updated without retraining the other, which is useful when extending the workflow to new chemical environments or larger system sizes. Extension of mlGW to one- and two-dimensional materials and to elements beyond the present H/C/N/O/F/Si set is currently underway, and the results will be reported elsewhere.

## Methods

### DFT and *GW* quasiparticle calculations

DFT calculations are carried out using QUANTUM ESPRESSO[58]. The Perdew-Burke-Ernzerhof (PBE) exchange-correlation functional[59] is used to account for the exchange-correlation effects. We use the optimized norm-conserving Vanderbilt pseudopotentials for all DFT calculations[60,61]. The Kohn-Sham wave functions were expanded in a plane-wave basis set with a cutoff energy of 90 Ry. Atomic-orbital descriptors were constructed using HPRO[62]. The self-consistent Kohn-Sham Hamiltonian obtained from QUANTUM ESPRESSO was represented in a nonorthogonal basis of SIESTA[63,64] numerical atomic orbitals with DZP quality. The Hamiltonian and overlap matrices were evaluated at each sampled wavevector, and the expansion coefficients were obtained by solving Eq. (2).

The quasiparticle calculations are carried out with the one-shot *GW* approximation ($G_0W_0$ approximation) implemented in the BerkeleyGW package[42]. We used the Hybertsen-Louie generalized plasmon pole model[11] to treat the frequency-dependent dielectric function and computed the dielectric matrices with a polarizability cutoff of 12 Ry. To balance accuracy and computational cost, the number of bands was set to ten times the number of valence bands to calculate the quasiparticle properties.

### The mlGW model and the training process

The initial node embedding features are projected into a 128-dimensional latent space through a linear layer followed by SiLU activation, layer normalization, and dropout. The resulting representations are subsequently updated through four residual message-passing blocks[38]. Atomic neighbors are defined using a cutoff radius of 5 Å, and interatomic distances are expanded using 32 radial basis functions. The Kohn-Sham energy of a target state relative to the VBM or HOMO energy is mapped into a 16-dimensional embedding through a learnable projection layer. The model was trained using the AdamW optimizer with an initial learning rate of $10^{-3}$, weight decay of $10^{-5}$, and batch size of 128. A smooth-L1 loss with a transition parameter of 0.03 was used as the training objective.

### Training datasets for machine-learned Hamiltonians

Separate machine-learned Hamiltonian models were trained for diamond and silicon using DFT Hamiltonian data generated from 16- and 54-atom bulk supercells.

Each dataset contained 1,100 structures, comprising 700 configurations sampled from *ab initio* molecular dynamics (AIMD) simulations and 400 randomly perturbed configurations. The AIMD simulations covered temperatures from 100 to 700 K in increments of 100 K, with 50 snapshots collected at each temperature for each supercell size. For each material, the dataset was divided into 880 training, 110 validation, and 110 test structures.

A separate machine-learned Hamiltonian model was trained from scratch for hydrogen-passivated Si nanoclusters using 596 Si-based structures, comprising 496 hydrogen-passivated nanoclusters, 80 hydrogen-passivated slabs, and 20 bulk Si configurations. These structures were divided into 477 training, 60 validation, and 59 test configurations. The training nanoclusters contained 14 to 180 atoms, and gauge standardization was enabled during training. Larger nanoclusters used in subsequent quasiparticle calculations were excluded from the machine-learned Hamiltonian training dataset. Electronic eigenvalues and eigenvectors were obtained by diagonalizing machine-learned Hamiltonians using the ScaLAPACK[37] package.

**Finite-temperature electron-phonon renormalization of band gaps**

Representative supercell structures to account for finite-temperature effects were generated using the one-shot method within the harmonic approximation proposed by Giustino and coworkers[48]. For a supercell containing *N* atoms, the atomic displacement is given by

$$\Delta\tau_{\kappa}(T) = \frac{1}{\sqrt{M_{\kappa}}} \sum_{\nu=1}^{3(N-1)} (-1)^{\nu-1} \varepsilon_{\kappa\nu} \sigma_{\nu,T}\,, \tag{6}$$

where $M_{\kappa}$ is the atomic mass, $\varepsilon_{\kappa\nu}$ is the phonon eigenvector of mode $\nu$, $\omega_{\nu}$ is the corresponding phonon angular frequency, and

$$\sigma_{\nu,T} = \sqrt{\left(2n_{\nu,T}+1\right)\frac{\hbar}{2\omega_{\nu}}}, \tag{7}$$

$$n_{\nu,T} = \left[\exp\left(\frac{\hbar\omega_{\nu}}{k_B T}\right) - 1\right]^{-1}. \tag{8}$$

The phonon modes are ordered by increasing frequency, and the alternating signs are used to construct a single representative configuration that approaches the approaches the thermally averaged result in the large-supercell limit. In this work, we generated 9×9×9 supercells for both diamond and silicon at each temperature.

**Data availability:** The dataset supporting the findings of this work, together with the core code required to reproduce the reported results and the trained model weights, is openly available at https://github.com/fanzhang519-hue/ML-GW-correction.

**Acknowledgements**: We acknowledge financial support from the National Natural Science Foundation of China (12474221, 12534012), the Natural Science Foundation of Jiangsu Province (BK20252000), and the Fundamental Research Funds for the Central Universities (020414380192).

**References:**

1 Chandrasekaran, A. *et al.* Solving the electronic structure problem with machine learning. *npj Comput. Mater.* **5**, 22 (2019).

2 Bartók, A. P., Payne, M. C., Kondor, R. & Csányi, G. Gaussian Approximation Potentials: The Accuracy of Quantum Mechanics, without the Electrons. *Phys. Rev. Lett.* **104**, 136403 (2010).

3 Behler, J. & Parrinello, M. Generalized neural-network representation of high-dimensional potential-energy surfaces. *Phys. Rev. Lett.* **98**, 146401 (2007).

4 Li, H. *et al.* Deep-learning density functional theory Hamiltonian for efficient ab initio electronic-structure calculation. *Nat. Comput. Sci.* **2**, 367 (2022).

5 Gu, Q. *et al.* Deep learning tight-binding approach for large-scale electronic simulations at finite temperatures with ab initio accuracy. *Nat. Commun.* **15**, 6772 (2024).

6 Zhong, Y., Yu, H., Su, M., Gong, X. & Xiang, H. Transferable equivariant graph neural networks for the Hamiltonians of molecules and solids. *npj Comput. Mater.* **9**, 182 (2023).

7 Batzner, S. *et al.* E(3)-equivariant graph neural networks for data-efficient and accurate interatomic potentials. *Nat. Commun.* **13**, 2453 (2022).

8 Fiedler, L. *et al.* Predicting electronic structures at any length scale with machine learning. *npj Comput. Mater.* **9**, 115 (2023).

9 Zhang, L., Han, J., Wang, H., Car, R. & E, W. Deep Potential Molecular Dynamics: A Scalable Model with the Accuracy of Quantum Mechanics. *Phys. Rev. Lett.* **120**, 143001 (2018).

10 Kohn, W. & Sham, L. J. Self-consistent equations including exchange and correlation effects. *Phys. Rev.* **140**, A1133 (1965).

11 Hybertsen, M. S. & Louie, S. G. Electron correlation in semiconductors and insulators:

Band gaps and quasiparticle energies. *Phys. Rev. B* **34**, 5390 (1986).
12 Golze, D., Dvorak, M. & Rinke, P. The GW Compendium: A Practical Guide to Theoretical Photoemission Spectroscopy. *Front. Chem.* **7**, 377 (2019).
13 Hedin, L. New method for calculating the one-particle Green's function with application to the electron-gas problem. *Phys. Rev.* **139**, A796 (1965).
14 Aryasetiawan, F. & Gunnarsson, O. The GW method. *Reports on Progress in Physics* **61**, 237 (1998).
15 Vetsch, N. *et al.* in *Proceedings of the International Conference for High Performance Computing, Networking, Storage and Analysis* 1–13 (Association for Computing Machinery, 2025).
16 Zhou, Z. *et al.* Low-rank inversion for single-node 10,000-atom plane-wave GW calculations. *Phys. Rev. Lett.* **137**, 156401 (2026).
17 Liu, P., Kaltak, M., Klimeš, J. & Kresse, G. Cubic scaling GW: Towards fast quasiparticle calculations. *Phys. Rev. B* **94**, 165109 (2016).
18 Wilhelm, J., Golze, D., Talirz, L., Hutter, J. & Pignedoli, C. A. Toward GW Calculations on Thousands of Atoms. *J. Phys. Chem. Lett.* **9**, 306 (2018).
19 Del Ben, M. *et al.* Accelerating Large-Scale Excited-State GW Calculations on Leadership HPC Systems. in SC20: International Conference for High Performance Computing, Networking, Storage and Analysis, 1-11 (2020).
20 Yu, V. W.-z. & Govoni, M. GPU Acceleration of Large-Scale Full-Frequency GW Calculations. *J. Chem. Theory Comput.* **18**, 4690 (2022).
21 Gao, W., Tang, Z., Zhao, J. & Chelikowsky, J. R. Efficient Full-Frequency GW Calculations Using a Lanczos Method. *Phys. Rev. Lett.* **132**, 126402 (2024).
22 Wu, W. *et al.* Enabling 13K-Atom Excited-State GW Calculations via Low-Rank Approximations and HPC on the New Sunway Supercomputer. in SC24: International Conference for High Performance Computing, Networking, Storage and Analysis, 1-14 (2024).
23 Kutepov, A. L. Self-consistent GW method: O(N) algorithm for polarizability and self energy. *Comput. Phys. Commun.* **257**, 107502 (2020).
24 Deuschle, L. *et al.* Towards Exascale Simulations of Nanoelectronic Devices in the GW Approximation. in SC24: International Conference for High Performance Computing, Networking, Storage and Analysis, 1 (2024).
25 Zhang, M.-Y., Lin, P., Shi, R. & Ren, X. Low-Scaling GW Calculations of Quasi-Particle Energies for Extended Systems within the Numerical Atomic Orbital Framework. *J. Chem. Theory Comput.* **22**, 5770 (2026).
26 Çaylak, O. & Baumeier, B. Machine Learning of Quasiparticle Energies in Molecules and Clusters. *J. Chem. Theory Comput.* **17**, 4891 (2021).
27 Knøsgaard, N. R. & Thygesen, K. S. Representing individual electronic states for machine learning GW band structures of 2D materials. *Nat. Commun.* **13**, 468 (2022).
28 Pant, N., Ha, V.-A., Kim, D. & Giustino, F. Electron correlation in semiconductors and insulators via symbolic regression. *Phys. Rev. B* **114**, 105112 (2026).
29 Hou, B., Wu, J. & Qiu, D. Y. Unsupervised representation learning of Kohn–Sham states and consequences for downstream predictions of many-body effects. *Nat.*

*Commun.* **15**, 9481 (2024).

30 Hou, B., Xu, X., Wu, J. & Qiu, D. Y. MBFormer: A General Transformer-based Learning Paradigm for Many-body Interactions in Real Materials. *arXiv:2507.05480* (2025).

31 Abdelghany, R. A., Hsu, C.-E., Hsueh, H.-C., Tsai, Y.-H. & Chung, M.-C. Machine learning framework for GW calculations across molecular dynamics trajectories. *Machine Learning: Science and Technology* **7**, 045003 (2026).

32 Venturella, C. *et al.* Unified deep learning framework for many-body quantum chemistry via Green's functions. *Nat. Comput. Sci.* **5**, 502 (2025).

33 Shen, W. *et al.* Learning the self-energy through equivariant neural networks. *arXiv:2610.23717* (2026).

34 Fediai, A., Reiser, P., Peña, J. E. O., Wenzel, W. & Friederich, P. Interpretable delta-learning of GW quasiparticle energies from GGA-DFT. *Machine Learning: Science and Technology* **4**, 035045 (2023).

35 Baum, D., Förster, A. & Visscher, L. Transfer learning of GW Bethe–Salpeter equation excitation energies. *Chem. Sci.* **17**, 8090 (2026).

36 Qian, C., Vitartas, V., Kermode, J. R. & Maurer, R. J. Equivariant electronic Hamiltonian prediction with many-body message passing. *npj Comput. Mater.* **12**, 169 (2026).

37 Blackford, L. S. *et al. ScaLAPACK users' guide*. (SIAM, 1997).

38 Hu, W. *et al.* Strategies for pre-training graph neural networks. *arXiv:1905.12265* (2019).

39 Fediai, A., Reiser, P., Peña, J. E. O., Friederich, P. & Wenzel, W. Accurate GW frontier orbital energies of 134 kilo molecules. *Scientific Data* **10**, 581 (2023).

40 Jain, A. *et al.* Commentary: The Materials Project: A materials genome approach to accelerating materials innovation. *APL Materials* **1**, 011002 (2013).

41 Horton, M. K. *et al.* Accelerated data-driven materials science with the Materials Project. *Nat. Mater.* **24**, 1522 (2025).

42 Deslippe, J. *et al.* BerkeleyGW: A massively parallel computer package for the calculation of the quasiparticle and optical properties of materials and nanostructures. *Comput. Phys. Commun.* **183**, 1269 (2012).

43 Karsai, F., Engel, M., Flage-Larsen, E. & Kresse, G. Electron–phonon coupling in semiconductors within the GW approximation. *New Journal of Physics* **20**, 123008 (2018).

44 Lautenschlager, P., Allen, P. B. & Cardona, M. Temperature dependence of band gaps in Si and Ge. *Phys. Rev. B* **31**, 2163 (1985).

45 Zollner, S., Gopalan, S. & Cardona, M. The temperature dependence of the band gaps in InP, InAs, InSb, and GaSb. *Solid State Commun.* **77**, 485 (1991).

46 Capaz, R. B., Spataru, C. D., Tangney, P., Cohen, M. L. & Louie, S. G. Temperature Dependence of the Band Gap of Semiconducting Carbon Nanotubes. *Phys. Rev. Lett.* **94**, 036801 (2005).

47 Gong, X., Li, Z. & Louie, S. G. Large Many-Electron Effects in the Temperature-Dependent Electron-Phonon Renormalization of Semiconductor Band Gaps. *Phys. Rev.*

*Lett.* **136**, 216401 (2026).
48 Zacharias, M. & Giustino, F. One-shot calculation of temperature-dependent optical spectra and phonon-induced band-gap renormalization. *Phys. Rev. B* **94**, 075125 (2016).
49 Miglio, A. *et al.* Predominance of non-adiabatic effects in zero-point renormalization of the electronic band gap. *npj Comput. Mater.* **6**, 167 (2020).
50 Monserrat, B., Conduit, G. J. & Needs, R. J. Extracting semiconductor band gap zero-point corrections from experimental data. *Phys. Rev. B* **90**, 184302 (2014).
51 Cardona, M. & Thewalt, M. L. W. Isotope effects on the optical spectra of semiconductors. *Reviews of Modern Physics* **77**, 1173 (2005).
52 O'Donnell, K. P. & Chen, X. Temperature dependence of semiconductor band gaps. *Appl. Phys. Lett.* **58**, 2924 (1991).
53 Alex, V., Finkbeiner, S. & Weber, J. Temperature dependence of the indirect energy gap in crystalline silicon. *J. Appl. Phys.* **79**, 6943 (1996).
54 Thomas, P. S. *et al.* StochasticGW-GPU: Rapid Quasi-Particle Energies for Molecules beyond 10,000 Atoms. *J. Chem. Theory Comput.* **22**, 3960 (2026).
55 Wolf, O. *et al.* Doping and Quantum Confinement Effects in Single Si Nanocrystals Observed by Scanning Tunneling Spectroscopy. *Nano Lett.* **13**, 2516 (2013).
56 Doherty, M. W. *et al.* The nitrogen-vacancy colour centre in diamond. *Phys. Rep.* **528**, 1 (2013).
57 Gong, X. *et al.* Deep Learning GW Quasiparticle Hamiltonians for Many-Body Excited-State Electronic Structure at Scale. *arXiv:2609.36962* (2026).
58 Giannozzi, P. *et al.* QUANTUM ESPRESSO: a modular and open-source software project for quantum simulations of materials. *J. Phys.: Condens. Matter* **21**, 395502 (2009).
59 Perdew, J. P., Burke, K. & Ernzerhof, M. Generalized Gradient Approximation Made Simple. *Phys. Rev. Lett.* **77**, 3865 (1996).
60 Hamann, D. R. Optimized norm-conserving Vanderbilt pseudopotentials. *Phys. Rev. B* **88**, 085117 (2013).
61 van Setten, M. J. *et al.* The PseudoDojo: Training and grading a 85 element optimized norm-conserving pseudopotential table. *Comput. Phys. Commun.* **226**, 39 (2018).
62 Gong, X., Louie, S. G., Duan, W. & Xu, Y. Generalizing deep learning electronic structure calculation to the plane-wave basis. *Nat. Comput. Sci.* **4**, 752 (2024).
63 Soler, J. M. *et al.* The SIESTA method for ab initio order-N materials simulation. *J. Phys.: Condens. Matter* **14**, 2745 (2002).
64 García, A. *et al.* Siesta: Recent developments and applications. *J. Chem. Phys.* **152**, 204108 (2020).